\documentclass[11pt,twoside]{article}
\usepackage{asp2010,natbib}
\resetcounters

\begin{document}

\title{Snooping around the big dog: VY\,CMa as seen with {\it Herschel}/HIFI}
\author{E. De Beck$^1$, L. Decin$^{1,2}$, K. M. Menten$^3$, A. Marston$^4$, D. Teyssier$^4$, and the HIFISTARS team
\affil{$^1$Department of Physics and Astronomy, Institute for  Astronomy, K.U.Leuven, Celestijnenlaan 200D, B-3001 Leuven, Belgium}
\affil{$^2$Astronomical Institute ``Anton Pannekoek'', University of Amsterdam, Science Park XH, Amsterdam, The Netherlands}
\affil{$^3$Max-Planck-Institut f\"ur Radioastronomie, Auf dem H\"ugel 69, D-53121 Bonn, Germany}
\affil{$^4$European Space Astronomy Centre, ESA, P.O. Box 78, E-28691 Villanueva de la Ca\~nada, Madrid, Spain}}

\begin{abstract} In the framework of the HIFISTARS guaranteed time key programme, we measured more than 70 molecular emission lines with high signal-to-noise ratio towards VY\,CMa using the high-resolution HIFI spectrometer \citep{degraauw2010} on board the \emph{Herschel}\footnote{{\it Herschel} is an ESA space observatory with science instruments provided by European-led Principal Investigator consortia and with important participation from NASA.} satellite. The kinematic information obtained from the measured water lines supports the hypothesis of multiple outflow components. The observed high-intensity maser lines give no indication for strong polarisation.
\end{abstract}

\section{A simple intro to a complex outflow}
VY CMa is a red supergiant at a distance of 1.1\,kpc, with a luminosity of $3\times10^5$\,L$_{\odot}$, T$_{\mathrm{eff}}$ $\sim$2800\,K, and a mass $\sim$25\,M$_{\odot}$. \cite{decin2006} derived a gas mass-loss rate of $\sim$$3.2\times10^{-4}$\,M$_{\odot}$/yr. Millimeter interferometry reveals three kinematic components: \emph{(1)} a dense, compact, and dusty central component, embedded in \emph{(2)} a more diffuse and extended envelope, and \emph{(3)} a high-velocity bipolar outflow \citep{muller2007}. 

\section{Watery outflows}
All H$_2$O lines in the left panel of Fig.\,1, have very similar shapes and match a Gaussian profile centred at $\varv_{\mathrm{LSR}}=21$\,km/s, with FWHM equal to the terminal velocity $\varv_{\mathrm{e}}$ derived from ground-based CO data, i.e. 46.5\,km/s. Even higher velocities, up to 60\,km/s, may be reached in the blue wing of the water lines (not shown in Fig.\,1). Compared to a Gaussian we see strong self-absorption in the blue, excess emission in the red, and a "shoulder" around 10\,km/s. The small bumps at the extreme blue velocities are the contributions from the outer layers moving toward the observer. This is as expected for the optically thick case. The thermal water emission thus likely traces all three outflow components.

\section{Strong maser lines}
Four strong maser lines were detected with HIFI: two rotational transitions of SiO in the first vibrationally excited state (13-12,v=1) and (15-14,v=1), and two of H$_2$O, both in the ground-vibrational state. \cite{neufeld1991} predicted p-H$_2$O($5_{2,4}-4_{3,1}$) and o-H$_2$O($5_{3,2}-4_{4,1}$) to be maser lines. Their respective observed peak strengths are 3.4\,K and 1.8\,K in main beam temperature. It is remarkable that the o-H$_2$O maser was observed twice, with an interval of three weeks, and no variation was detected on this timescale \citep{harwit2010}. Apart from the Goldreich-Kylafis effect in the wings of the o-H$_2$O maser, mentioned by \cite{harwit2010}, we did not detect any strong polarisation effects for any of these maser lines.

\begin{figure}\centering
\includegraphics[height=5.8cm]{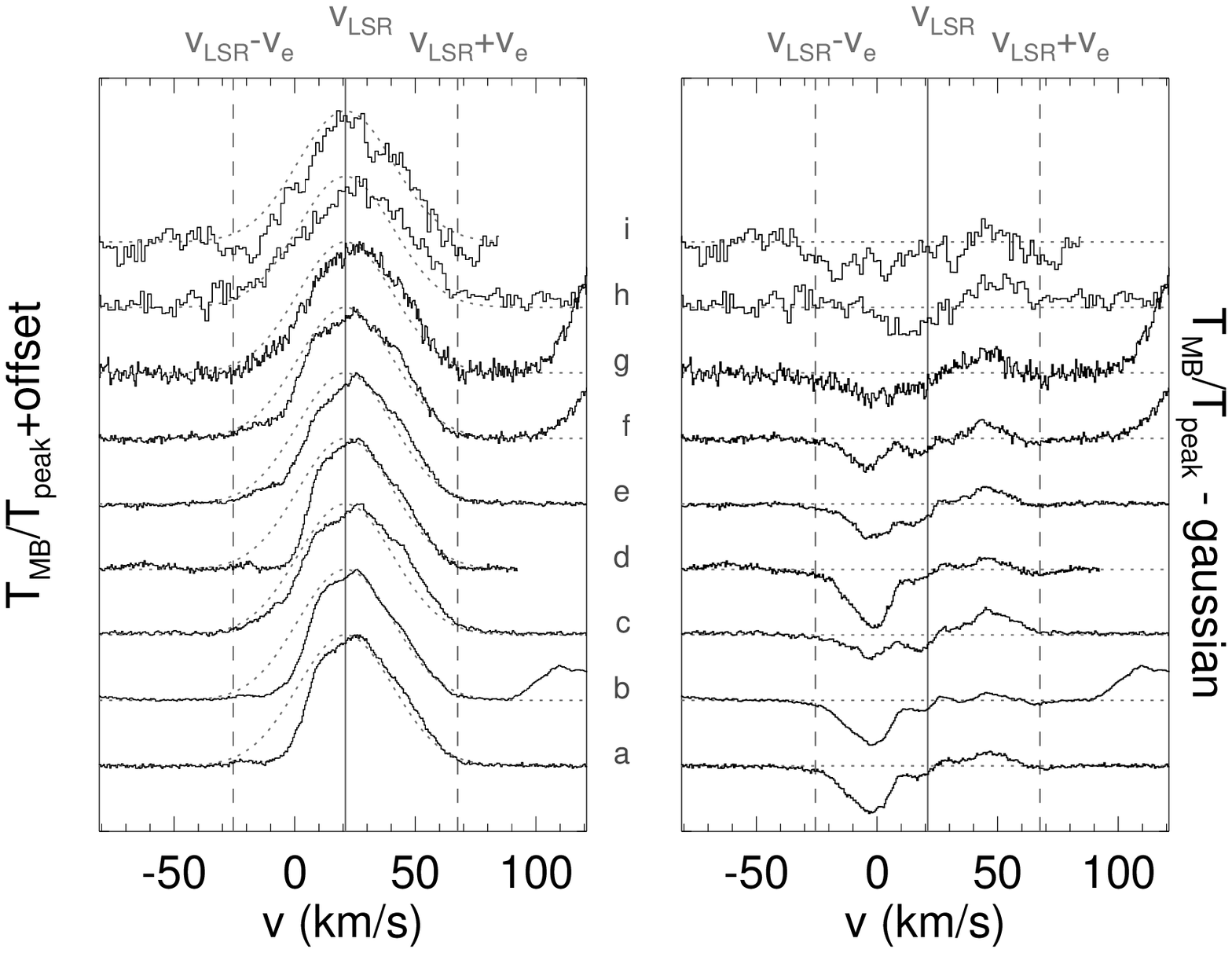}
\includegraphics[height=5.8cm]{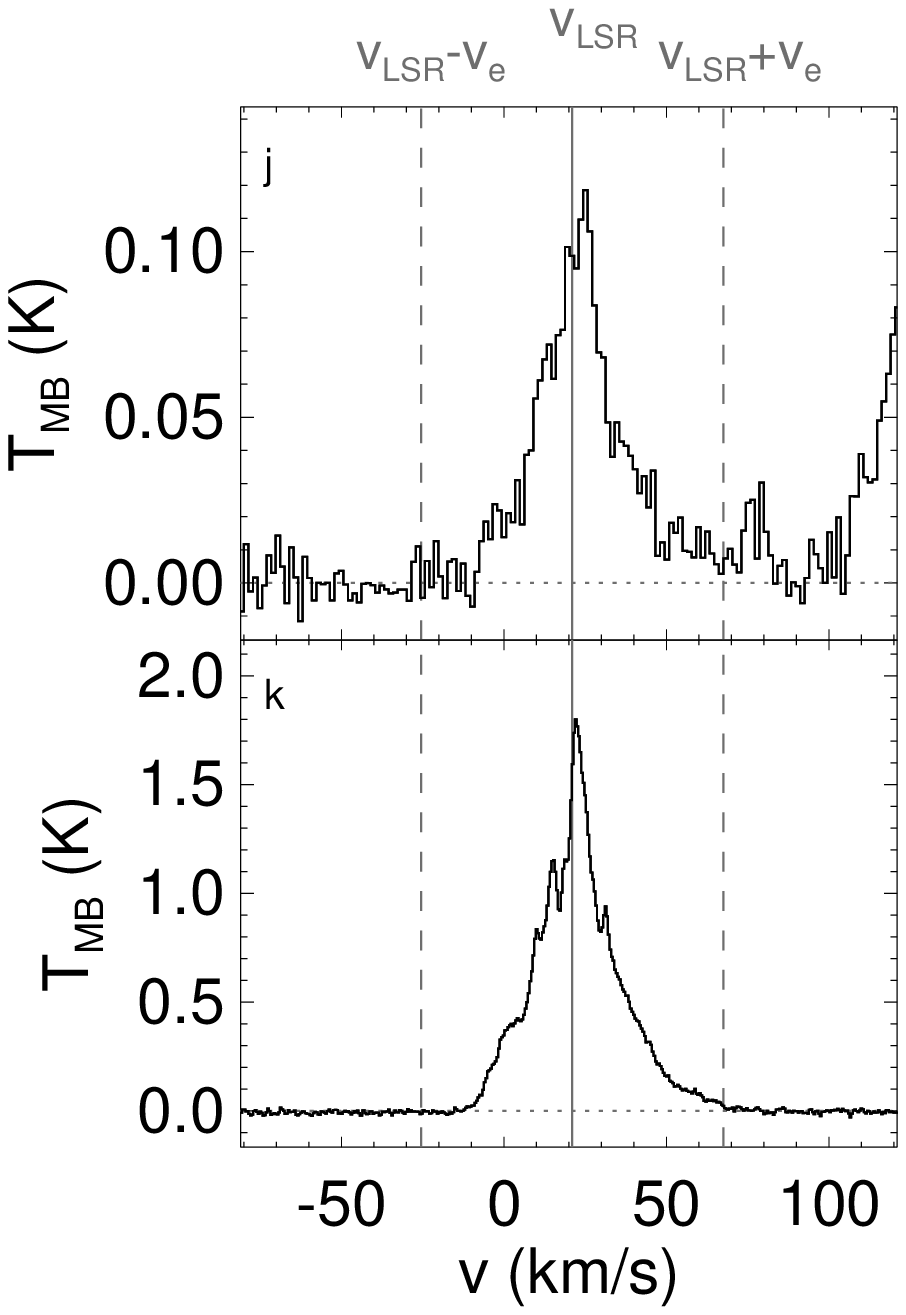}
\caption{\emph{Left ---} comparing the shape of H$_2$O lines: 
\emph{(a)} p-H$_{2}$O(1$_{1,1}$-0$_{0,0}$),   
\emph{(b)} o-H$_{2}$O(1$_{1,0}$-1$_{0,1}$),  
\emph{(c)} p-H$_{2}$O(2$_{1,1}$-2$_{0,2}$),  
\emph{(d)} o-H$_{2}$O(3$_{0,3}$-2$_{1,2}$),  
\emph{(e)} o-H$_{2}$O(3$_{1,2}$-3$_{0,3}$),  
\emph{(f)} o-H$_{2}$O(3$_{1,2}$-2$_{2,1}$),  
\emph{(g)} o-H$_{2}$O(3$_{2,1}$-3$_{1,2}$),  
\emph{(h)} p-H$_{2}$O(4$_{2,2}$-4$_{1,3}$),  
\emph{(i)} o-H$_{2}$O(5$_{3,2}$-5$_{2,3}$).  
The dashed profiles are Gaussian as described in the text. 
\emph{Middle ---} Gaussian-subtracted profiles.
\emph{Right ---} selected maser lines: 
\emph{(j)} SiO(15-14,v=1),  
\emph{(k)} o-H$_{2}$O(5$_{3,2}$-4$_{4,1}$). 
The vertical dashed lines indicate the region $[\varv_{\mathrm{LSR}}-\varv_{\mathrm{e}};\varv_{\mathrm{LSR}}+\varv_{\mathrm{e}}]$.
}
\end{figure}

\acknowledgements 
E. De Beck and L. Decin wish to acknowledge the financial support by FWO under grant number G.0470.07.

\bibliography{DeBeck}

\begin{thebibliography}{}
\expandafter\ifx\csname natexlab\endcsname\relax\def\natexlab#1{#1}\fi
\expandafter\ifx\csname url\endcsname\relax
  \def\url#1{\texttt{#1}}\fi
\expandafter\ifx\csname urlprefix\endcsname\relax\def\urlprefix{URL }\fi
\providecommand{\eprint}[2][]{\url{#2}}

\bibitem[{{de Graauw} et~al.(2010){de Graauw}, {Helmich}, {Phillips}, \&
  {others}}]{degraauw2010}
{de Graauw}, T., {Helmich}, F.~P., {Phillips}, T.~G., \& {others} 2010, \aap,
  518, L6+

\bibitem[{{Decin} et~al.(2006){Decin}, {Hony}, {de Koter} et~al.}]{decin2006}
{Decin}, L., {Hony}, S., {de Koter}, A., et~al. 2006, \aap, 456, 549.
  \eprint{arXiv:astro-ph/0606299}

\bibitem[{{Harwit} et~al.(2010){Harwit}, {Houde}, {Sonnentrucker}
  et~al.}]{harwit2010}
{Harwit}, M., {Houde}, M., {Sonnentrucker}, P., et~al. 2010, \aap, 521, L51+.
  \eprint{1007.0905}

\bibitem[{{Muller} et~al.(2007){Muller}, {Dinh-V-Trung}, {Lim}, {Hirano},
  {Muthu}, \& {Kwok}}]{muller2007}
{Muller}, S., {Dinh-V-Trung}, {Lim}, J., {Hirano}, N., {Muthu}, C., \& {Kwok},
  S. 2007, \apj, 656, 1109

\bibitem[{{Neufeld} \& {Melnick}(1991)}]{neufeld1991}
{Neufeld}, D.~A., \& {Melnick}, G.~J. 1991, \apj, 368, 215

\end{thebibliography}

\end{document}